\documentclass[12pt]{article}  
\usepackage{amssymb}
\def\R{{\mathbb R}}
\def\M{{\mathbb M}}
\def\P{{\mathbb P}}
\def\Z{{\mathbb Z}}
\def\Cl{{\mathbb C}\ell}
\def\be{\begin{equation}}
\def\ee{\end{equation}}
\def\bea{\begin{eqnarray}}
\def\eea{\end{eqnarray}}
\begin{document}
\title{On Conformally Compactified Phase  Space}  
\author{P. Budinich}
\maketitle
\begin{abstract}
 Conformally compactified phase space  is conceived as an automorphism  space
for the global action of the extended conformal group. Space time and momentum
space appear then as conformally dual, that is conjugate with respect to
conformal reflections. If now the former, as generally agreed, is appropriate
for the description of classical mechanics in euclidean geometrical form,
then the latter results appropriate for the description of quantum mechanics
in spinor geometrical form. In such description, fermion multiplets will
naturally appear as consequence of higher symmetries and furthermore, the
euclidean geometry, bilinearly resulting from that of spinors, will a priori
guarantee the absence of ultraviolet divergences when dealing with quantum field
theories. Some further possible consequences of conformal reflections of interest 
for physics, are briefly outlined.
\end{abstract}
\section{Introduction}

\noindent The discovery of Maxwell's equations covariance with respect to the
conformal group $C = \left\{ L,  D,  P_4, S_4 \right\} $ 
(where $L$,$D$,$P_4$,$S_4$ mean \mbox{Lorentz-,} \mbox{Dilatation-,} \mbox{
Poincar\`e-,} and special conformal transformations respectively building up
$C$)~\cite{1} have induced several  
authors~\cite{2} to conjecture that Minkowski space-time $\M=\R^{3,1}$ may be densely 
contained in conformally compactified\footnote{From which 
Robertson-Walker space-time $\M_{RW}= S^3 \times R^1$ is obtained,
familiar to cosmologists (being at the origin of the Cosmological
Principle~\cite{3}), 
where $R^1$ is conceived as the infinite covering of $S^1$.} space-time $\M_c$:
\be
 \M_c=\frac{S^3 \times S^1}{\Z_2} \label{eq1}
\ee
often conceived as the homogeneous space of the conformal group:
$$
\M_c=\frac{C}{c_1}
$$ 
where $c_1 = \left\{ L, D,  S_4 \right\}$ is the stability group of the origin $x_\mu=0$.
 
As well known $C$ may be linearly represented by $SO(4,2)$, acting in $\R^{4,2}$,
containing the Lorentz group $SO(3,1)$ as a subgroup which, because of the
relevance of space-time reflections for natural phenomena, should be
extended to $O(3,1)$. But then $SO(4,2)$ should be also extended to $O(4,2)$
including conformal reflections $I$ (~with respect to hyperplanes orthogonal
to the $5^{th}$ and $6^{th}$ axis~), whose relevance for physics should then be
expected as well. To start with, in fact, in this case $\M_c=C/c_1$ seems not
to be the only automorphism space of $O(4,2)$, since:
\be
 I \M_c I^{-1} = I \frac{C}{c_1} I^{-1} = \frac{C}{c_2} = \P_c = \frac{S^3
\times S^1}{\Z_2} \label{eq2}
\ee
where $c_2 =\left\{ L, D, P_4\right\}$ is the stability group of  
infinity.  Therefore, being $c_1$ and $c_2$ conjugate, $\M_c$ and $\P_c$ are two copies of
the same homogeneous space ${\mathcal H}$ of the conformal group including
reflections, and, as we will see, both are needed to represent the group linearly .
Because of eq. (\ref{eq2}) we will call $\M_c$ and $\P_c$  conformally dual.

There are good arguments~\cite{4} (see also footnote \ref{fn1}) in favor of the
hypothesis that $\P_c$ may represent conformally compactified momentum space
$\P=\R^{3,1}$. In this case $\M_c$ and $\P_c$ build up conformally compactified phase
space, which then is a space of automorphism of the conformal group
$C$ including reflections.

\section{Conformally compactified phase space}

For simultaneous compactification of $\M$ and $\P$ in $M_c$ and $\P_c$ no exact Fourier
transform is known. It can be only approximated by a finite lattice phase
space~\cite{5}.

An exact Fourier transform may be defined instead in the 2-dimensional
space time when $\M =\R^{1,1}=\P$ and for which
\be
  \M_c=\frac{S^1 \times S^1}{\Z_2}=\P_c \label{eq4}
\ee
Then inscribing in each $S_1$, of radius $R$, of $\M_c$ and in each $S_1$ of radius $K$
of $\P_c$ a regular polygon with
\be
 2 N = 2 \pi R K \label{eq5}
\ee
vertices any function $f\left(x_{mn}\right)$ defined in the resulting lattice 
$M_L \subset \M_c$ is correlated to the Fourier-transformed 
$ F\left( k_{\rho\tau} \right)$ on the 
$P_L \subset \P_c$ lattice by the finite Fourier series:
\bea
 f\left(x_{nm}\right)=\frac{1}{2 \pi R^2} \sum_{\rho, \tau = -N}^{N-1} 
  \varepsilon^{\left(n \rho - m \tau\right)} F\left(k_{\rho \tau}\right) \nonumber \\
\;\;\;\;\; \label{eq6} \\
 F\left( k_{\rho\tau} \right) = \frac{1}{2 \pi K^2} \sum_{n,m = -N}^{N-1}
  \varepsilon^{-\left(n \rho - m \tau \right)} f \left(x_{nm}\right) \nonumber
\eea
where $ \varepsilon=e^{i \frac{\pi}{N}}$ is the $2N$-root of unit. 
They may be called Fourier transforms since either for $R \rightarrow \infty$
or $K  \rightarrow \infty$, (or both) they coincide with the standard
ones. This further confirms the identification of $\P_c$ with momentum
space which is here on purpose characterized geometrically rather then
algebraically~(Poisson bracket). On this model the action of the conformal
group $O(2, 2)$ may be easily operated and tested.

\section{Conformal duality}
The non linear, local action of $I$ on $\M$ is well known; for $x_\mu \in \M$ we have:
\be
 I \; :  \; x_\mu \rightarrow I\left( x_\mu \right) = \pm \frac{x_\mu}{x^2} \label{eq7}
\ee
For $x^2 \not= 0$ and $x_\mu$ space-like, if $x$ indicates the distance of a
point from the 
origin, we have \footnote{$I$ maps every point, at a distance $x$ from the
center, in the sphere $S^2$, to a point outside of it at a distance $x^{-1}$.
For $\M=\R^{2,1}$ the sphere $S^2$ reduces to a circle $S^1$ and then
(\ref{eq7p}) reminds Target Space duality in string theory~\cite{6}, which then
might be the consequence of conformal inversion. For $\M=\R^{1,1}$, that is
for the two dimensional model $I$ may be locally represented through quotient
rational transformation by means of $I=i \sigma_2$ and the result is
(\ref{eq7p}), which in turn represents the action of $I$ for the conformal
group $G= \left\{ D, P_1, T_1\right\}$ on a straight line $\R^1$.} :
\be
 I \; :  \; x \rightarrow I\left( x \right) =  \frac{1}{x} .
 \label{eq7p}
\ee
Since $\M$ is densely contained in $\M_c$, $x_\mu$ defines a point of the homogeneous
space ${\mathcal H}$, of automorphisms for $C$; as such $x_\mu$ must then be conceived as
dimensionless in (\ref{eq7}), as usually done in mathematics. 
Therefore for physical applications, to represent space-time we must
substitute $x_\mu$ with	$x_\mu/l$, where $l$ represents an (arbitrary) unit of
lengths, then from (\ref{eq7}) and (\ref{eq7p}) we obtain:

 \underline{ Proposition $P_1$}: Conformal reflections determine a map in 
space of the microworld  to the macroworld~( with respect to $l$) and vice-versa.

The conformal group may be well represented in momentum space $\P=\R^{3,1}$, densely
contained in $\P_c$, where the action of $I$ induces non linear transformation
like (\ref{eq7}) and (\ref{eq7p}), where $x_\mu$ and $x$ are substituted by 
$k_\mu \in \P$ and $k$. If we then take (\ref{eq7p}) and the corresponding for
$\P$ we obtain (see also footnote 2): 
\be
I \; :  \; x k \rightarrow I\left( x k \right) = 
\frac{1 }{x k} \label{eq8}
\ee
Now physical momentum $p$ is obtained after multiplying the wave-number $k$ by
an (arbitrary) unit of action $H$ by which (\ref{eq8}) becomes:
\be
I \; :  \; \frac{x p}{H} \rightarrow I\left( \frac{x p}{H} \right) = 
\frac{H }{x p} \label{eq9}
\ee
from which we obtain:

\underline{Proposition $P_2$}: Conformal reflections determine a map, 
in phase space of the
world of micro actions to the one of macro actions (with respect to $H$), and
vice-versa.

Now if we choose for the arbitrary unit $H$ the Planck's constant $\hbar$ then from
propositions $P_1$ and $P_2$ we have:

\underline{Corollary $C_2$}: Conformal reflections determine a map between classical and
quantum mechanics.

Let us now remind the identifications $\M_c \equiv C/c_1$ and $\P_c \equiv
C/c_2$ to be conceived as two copies of 
the homogeneous space ${\mathcal H}$, representing conformally compactified
space-time and 
momentum space respectively, and that $I \M_c I^{-1} =\P_c$ and then
$I$ represents a map\footnote{
The action of $I$ may be rigorously tested in the two dimensional model where
$I(x_{nm}) =k_{nm}$ and the action of $I$ is linear in the compactified phase
space at difference with its local non linear action in $\M$ and $\P$, as will be further
discussed elsewhere.} 
of every point $x_\mu$ of $\M$ to a point $k_\mu$ of
$\P$ : $ I \rightarrow I(x_\mu) = p_\mu$ and we have:

\underline{Proposition $P_3$}: Conformal reflections determine a map between
space-time and momentum space.

Let us now assume, as the history of celestial mechanics suggests, that
space-time $\M$ is the most appropriate for the description of classical
mechanics then, as a consequence of propositions $P_1$, $P_3$ and of corollary
$C_2$, momentum space should be the most appropriate for the description of
quantum mechanics. The legitimacy of this conjecture seems in fact to be
supported by spinor geometry as we will see.

\section{Quantum mechanics in momentum space}

	Notoriously the most elementary constituents of matter are fermions,
represented by spinors, whose geometry, as formulated by its discoverer E.
Cartan~\cite{7} has already the form of equations of motions for fermions in
momentum space.

 In fact given a pseudo euclidean, $2n$-dimensional vector space $V$ with
signature $\left(k,l \right)$; $k+l=2n$ and the associated Clifford algebra
$\Cl\left(k,l \right)$ with generators $\gamma_a$ a Dirac spinor $\psi$ is an 
element of the endomorphism space of $\Cl(k,l)$ and is defined by Cartan's equation
\be
 \gamma_a p^a \psi =0   \label{eq10} 
\ee
where $p_a$ are the components of a vector $p \in V$.

Now it may be shown that for the signatures $(k,l)= (3,1),(4,1)$,
the Weyl, (Maxwell), Majorana, Dirac equations, respectively may be
natural\-ly obta\-ined~\cite{8}, from (\ref{eq10}), precisely in momentum space. For the
signature $(4,2)$ eq. (\ref{eq10}) contains twistors equations and, for $\psi
\not= 0$, the vector  $p$ is null: $p_a p^a =0$ and the directions of $p$ form
the projective quadric\footnote{Since Weyl equation in $\P=\R^{3,1}$, out of
which Maxwell's equation for the electromagnetic tensor $F_{\mu\nu}$
(expressed bilinearly in term of Weyl spinors) may be obtained, is contained in
twistor equation in $\P=\R^{4,2}$, defining the projective quadric: $\P_c =
\left( S^3 \times S^1 \right)/\Z_2$. This is a further argument why momentum
space $\P$ should be densely contained in $\P_c$. \label{fn1}} $(S^3 \times S^1)/\Z_2$
identical to conformally compactified momentum space $\P_c$ given in (\ref{eq2}). 

For the signature $(5,3)$ instead one may easily obtain~\cite{8} from (\ref{eq10})
the equation
\be
\left( \gamma_\mu p^\mu + \vec{\pi} \cdot \vec{\sigma} \otimes \gamma_5+M\right) N =0
\label{eq11}
\ee
where 
$ \vec{\pi}=\langle \tilde{N} , \vec{\sigma} \otimes \gamma_5 N \rangle$ 
and $N=\left[ \matrix{ \psi_1 \cr \psi_2 } \right]$ ; 
 $M= \left[ \matrix { m & 0 \cr 0 & m} \right]$ ; 
 $\tilde N=\left[ \tilde \psi_1 , \tilde \psi_2 \right]$  ; 
 $\tilde \psi_j=\psi_j^\dagger \gamma_0$, 
with  $\psi_1$,$\psi_2$ - space-time Dirac spinors, and 
$\vec{\sigma}=\left( \sigma_1,\; \sigma_2,\;\sigma_3 \right)$ Pauli matrices.

	Eq. (\ref{eq11}) represents the equation, in momentum space, of the 
proton-neutron doublet interacting with the pseudoscalar pion triplet$
\vec{\pi}$.

	Also the equation for the electroweak model may be easily obtained~\cite{9} in
the frame of $\Cl(5,3)$.

	All this may further justify the conjecture that spinor geometry in
conformally compac\-tified momentum space is the appropriate arena for the
description of quantum mechanics of fermions.

It is remarkable that for signatures $(3,1)$, $(4,1)$ and $(5,3)$, $(7,1)$ (
while not for $(4,2)$~) the real
components  $p_a$ of $p$ may be bilinearly expressed in terms of spinors~\cite{8}.

	If spinor geometry in momentum space is at the origin of quantum mechanics
of fermions, then their multiplicity, observed in natural phenomena, could
be a natural consequence of the fact that a Dirac spinor associated with   
 $\Cl(2n)$ has $2^n$ components, which naturally splits in multiplets of space-time
spinors (as it already appears in eq. (\ref{eq11})). Since in this approach vectors
appear as bilinearly composed by spinors, some of the problematic aspects of
dimensional reduction could be avoided dealing merely with spinors~\cite{10}.

	Also rotations naturally arise, in spinor spaces as products of reflections
operated by the generators $\gamma_a$ of the Clifford algebras. These could then
be at the origin of the so called internal symmetry. This appears in eq.
(\ref{eq11}) where the isospin symmetry of nuclear forces arises from conformal
reflections which appear there as the units of the quaternion field of
numbers of which, proton-neutron equivalence, for strong interactions, could
be a realization in nature.
	For $\Cl(8)$ and higher Clifford algebras, and the associated spinors,
octonions could be expected to play a role as recently advocated~\cite{10}.

\section{Some further consequences of conformal duality}

	Compact phase space implies, for field theories, the absence of the
concept of infinity in both space-time and momentum and, provided we may
rigorously define Fourier dual manifolds, where fields may be defined, it
would imply also
the absence of both infrared and ultraviolet divergences in perturbation
expansions. This is, for the moment, possible only in the four-dimensional
phase space-model, where such manifold restricts to the lattice $M_L$ and $P_L$ 
on a double torus, where Fourier transforms (\ref{eq6}) hold. 
One could call  $M_L$ and $P_L$  the physical
spaces which are compact and discrete, to distinguish them from the
mathematical spaces $M_c$ and $P_c$ which are compact and continuous, the
latter are only conformally dual while the former are both conformally and
Fourier dual.

	In the realistic eight dimensional phase space one would also expect to
find physical spaces represented by lattices~\cite{5} as non commutative
geometry seems also to suggest~\cite{11}.

	Let us now consider as a canonical example of a quantum system:  the
hydrogen atom in stationary states. According to our hypothesis it should be
appropriated to deal with it in momentum space (as a non relativistic limit of
the Dirac equation for an electron subject to an external e.m. field). This
is possible as shown by v. Fock~\cite{12} through the integral equation:
\be
 \phi\left( p \right) = \frac{\lambda}{2 \pi^2} \int_{S^3} \, 
\frac{\phi\left(q\right)}{\left( p - q \right)^2 } d^3  q \label{eq12}
\ee
where $S^3$ is the one point compactification of momentum space $\P=\R^{3}$, 
and $\lambda=\frac{e^2}{\hbar c } \sqrt{\frac{p_0^2}{-2m E}}$, where 
$p_0$ is a unit of momentum and  $E$ the (negative) energy of the H-atom. 

For $\lambda = n+1$, $\phi(p) = {Y_{nlm}}{\left(\alpha\beta\gamma \right)}$ 
which are the harmonics on $S^3$, and for $p_0=mc$, we obtain:
$$
 E_n =- \frac{m e^4}{2 \hbar^2 \left( n+1 \right)^2}
$$
which are the energy eigenvalues of the H-atom.

	It is interesting to observe that eq. (\ref{eq12}) is a purely geometrical
equation, where the only quantum parameter\footnote{The geometrical
determination of this parameter in eq. (\ref{eq12}) (through harmonic
analysis, say) could furnish a clue to understanding the geometrical origin of
quantum mechanics. This was a persistent hope of the late Wolfgang Pauli.} 
is the dimensionless fine structure constant
$$
  \frac{e^2}{\hbar c} = \frac{1}{137,...}
$$

	According to this equation the stationary states of the H-atom may be
represented as eigenvibrations of the $S^3$ sphere (of radius $p_0$ ) in
conformally compactified momentum space and out of which the quanum numbers
$n$,$l$,$m$ result. If conformal duality is realized 
in nature then there should exist a classical system represented by
eigenvibration of $S^3$ in $\M_c$ or $\M_{RW}$.

	In fact this system could be the universe since recent observations on
distant galaxies (in the direction of N-S galactic poles) have revealed
that their distribution may be represented~\cite{13} by the $S^3$ eigenfunction
\be
 Y_{n,0,0} = k_n \frac{\rm{sin} \; \left(n+1\right) \chi }{\rm{sin} \; \chi} 
\label{eq13}
\ee
	with $k_n$ a constant, $\chi$ the geodetic distance from the center of the
corresponding eigenvibration on the $S^3$ sphere of the $\M_{RW}$ universe. 
Now $ Y_{n,0,0} $ is exactly equal to the eigenfunction of the H-atom however in
momentum space. If the astronomical observations will confirm eq. (\ref{eq13}) then the
universe and the H-atom would represent a realization in nature of
conformal duality. Here we have in fact that $ Y_{n,0,0}$  on $\P_c$
represents the (most symmetric) eigenfunction of the (quantum) H-atom and
the same $ Y_{n,0,0}$ on $\M_c$ may represent the (visible) mass distribution of the
(classical) universe. They could then be an example\footnote{There could be 
other examples of conformal duality represented by our
planetary system. In fact observe that in order to compare with the
density of matter the square of the $S^3$ harmonic $Y_{nlm}$ has to be taken. Now 
 $Y_{n00}^2$ presents maxima for $r_n=\left(n+\frac{1}{2} \right)r_0$, and
it has been shown by Y.K. Gulak~\cite{14} that the values of the large semi axes
of the 10 major solar planets satisfy this rule, which could then suggest
that they arise from a planetary cloud presenting the structure of a $S^3$
eigenvibrations, as will be discussed elsewhere.}  
of conformally dual systems: one classical and one quantum mechanical.

There could be another important consequence of conformal duality. in fact
suppose that $\M_c$ and $\P_c$ are also Fourier dual\footnote{
Even if, for
defining rigorously Fourier duality for $\M_c$ and $\P_c$ one may
have to abandon standard differential calculus, locally it may be assumed to
be approximately true, with reasonable confidence since the spacing of the
possible lattice will be extremely small. In fact taking for $K$  the Planck's
radius one could 
have of the order of $10^{30}$ points of the lattice per centimeter.
}. 
Then to the eigenfunctions: ${Y_{nlm}}{\left(\alpha\beta\gamma \right)}$ on
$\P_c$ of the H-atom there will correspond
on $\M=\R^{3,1}$ (densely contained in $\M_c$) their Fourier transforms; that is the
known eigenfunctions $\Psi_{nlm}\left(x_1,x_2,x_3 \right)$ in $\M=\R^{3,1}$  of 
the H-atom stationary states.

Now according to propositions $P_1$, $P_3$ and corollary $C_2$ for high values of the
action of the system; that is for high values of $n$ in  $Y_{nlm}$, the
system should be identified with the corresponding classical one, and in
space-time $\M$ where it is represented by $\Psi_{nlm}\left(x_1,x_2,x_3
\right)$.

In fact this is what postulated by the correspondence principle: for high
values of the quantum numbers the wavefunction $\Psi_{nlm}\left(x_1,x_2,x_3
\right)$
identifies with the Kepler orbits; that is the same system (with potential
proportional to $1/r$) dealt in the frame of classical mechanics. In this
way, at least in this particular example, the correspondence principle
appears as a consequence of conformal duality\footnote{Also the property of
Fourier transforms play a role. Consider in fact a classical system with fixed
orbits: a massive point particle on $S^1$, say. Its quantization appears in
the Fourier dual momentum space $Z$ : ($m=0,\pm1, \pm 2 ,\dots $) and for the
large $m$ the eigenfunctions identifies with $S^1$.} and precisely of propositions
$P_1$, $P_3$ and corollary $C_2$. Obviously at difference with the previous case of
duality, in this case it is the same system (the two body problem) dealt
once quantum mechanically in $\P_c$ and once classically in $\M_c$. What they keep
in common is the $SO(4)$ group of symmetry,  named accidental symmetry when
discovered by W. Pauli for the Kepler orbits, while here it derives from the
properties of conformal reflections, which preserve $S^3$, as seen from (\ref{eq2}).

\end{document}